\documentclass[twocolumn,aps,apl,floats,superscriptaddress]{revtex4-1}

\usepackage[american]{babel}
\usepackage{units}
\usepackage{textcomp} 
\usepackage{graphicx}
\usepackage{amsmath}
\usepackage{braket}
\usepackage{bm}
\usepackage{amssymb}
\usepackage{geometry}

\begin{document}

\renewcommand{\figurename}{Fig.}


\title{Gradient Index Metamaterial Based on Slot Elements}
\author{Oliver Paul$^{1}$, Benjamin Reinhard$^{1}$, Bernd Krolla$^{1}$, Ren\'{e} Beigang$^{1,2}$, and Marco Rahm$^{1,2}$}
\affiliation{$^1$Department of Physics and Research Center OPTIMAS, University of Kaiserslautern, Germany \\
\vspace{5 pt} $^2$Fraunhofer Institute for Physical Measurement Techniques IPM, Freiburg, Germany}


\date{\today}

\begin{abstract}
We present a gradient-index (GRIN) metamaterial based on an array of annular slots. The structure
allows a large variation of the effective refractive index under normal-to-plane incidence and thus
enables the construction of GRIN devices consisting of only a small number of functional layers.
Using full-wave simulations, we demonstrate the annular slot concept by means of a 3-unit-cell thin
GRIN lens for the terahertz (THz) range. In the presented realizations, we achieved an index
contrast of $\Delta n = 1.5$ resulting in a highly refractive lens suitable for focusing THz
radiation to a spot size smaller than the wavelength.
\end{abstract}

\maketitle


In the last ten years, metamaterials have emerged to be powerful tools for the manipulation of
light on the subwavelength scale. From the very first day of metamaterials, the scientific interest
has been driven by the possibility of guiding light by tailoring the effective material parameters.
In this respect, various concepts have been proposed, ranging from gradient index materials
\cite{smith2000,liu2009a,cheng2009}, to invisibility cloaks
\cite{liu2009b,cai2007,ruan2007,pendry2008} and other transformation optical structures
\cite{leonhardt2006,pendry2006,schurig2007,rahm2008,kundtz2010}. Most of the concepts have only
been experimentally realized in the microwave regime where comparatively simple fabrication
techniques allow a large freedom in the metamaterial design. However, when approaching higher
frequencies as the THz or the optical regime, the experimental realization becomes more challenging
since the standard fabrication methods, such as photo- or electron beam lithography, allow only the
fabrication of planar structures with a very limited number of layers. In order to compensate the
layer restriction in the high frequency range, it is therefore important to use metamaterials that
provide a high refractive index contrast. However, most of the metamaterial structures are very
lossy and allow only a moderate change of the refractive index within an acceptable transmission
window. In particular, this is the case for resonant elements which are usually associated with
high intrinsic losses. One possibility to create a non-resonant element is to choose the operation
frequency well below the resonance frequency and, hence, operate in the non-resonant region of the
structure. This approach has been applied in Ref.~$2-4$ where the achieved index contrast was
reported to be in the range of $\Delta n = 0.7- 0.9$. In this paper, we present an alternative
approach for a non-resonant, polarization-independent structure which allows a very large variation
of the refractive index of about $\Delta n = 1.5$ with high transmission and reasonable bandwidth
for practical applications.\\



\begin{figure}[t]
   \begin{center}
    \includegraphics[width=\columnwidth]{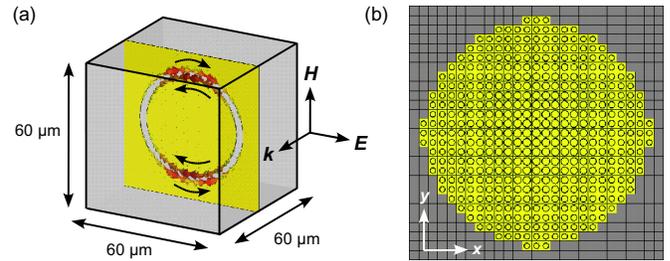}
    \end{center}
    \caption{(a) Unit cell of the annular slot structure with excited antisymmetric current mode. (b) Front view of the composed GRIN lens with a diameter of 25 unit cells.}
    \label{fig:unitcell}
\end{figure}

The unit cell of the proposed metamaterial consists of a $\unit[3]{\mbox{\textmu}m}$ wide annular
slot within a square metal patch as shown in Fig.~\ref{fig:unitcell}(a). The metal layer is
embedded in a cubic dielectric matrix with an edge length of \unit[60]{\mbox{\textmu}m}. The radius
of the annular slot was varied in order to alter the effective index of refraction. We analyzed the
electromagnetic properties of the metamaterial by full-wave time-domain simulations (CST Microwave
Studio). To consider a realistic model, the metal layer was implemented by $\unit[200]{nm}$ thick
copper with an electric conductivity of $\sigma = \unit[5.8 \times 10^7]{S/m}$. For the dielectric
matrix, we used benzocyclobutene (BCB) with a relative permittivity of~$\epsilon=2.67$ and a loss
tangent of $\tan \delta = 0.012$.

\begin{figure}[t]
   \begin{center}
    \includegraphics[width=\columnwidth]{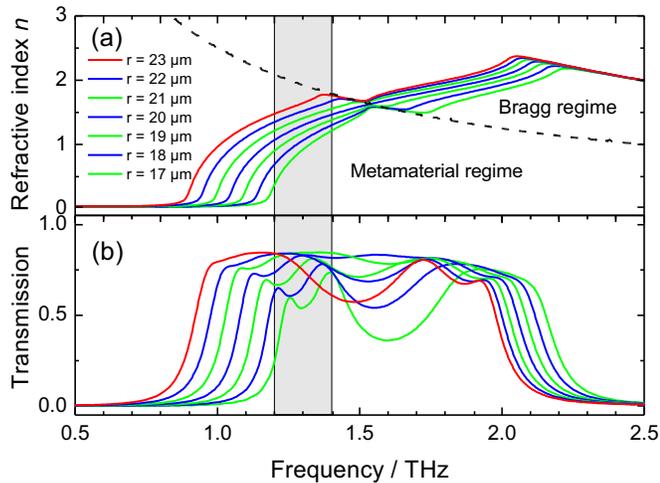}
    \end{center}
    \caption{(b) Real part of the effective index of refraction~$n$ for varying inner radius~$r$ of the slot. (b) Amplitude transmission for three unit cells in the direction of propagation. The spectral operation range of the GRIN lens is
indicated by a grey box.}
    \label{fig:transmission}
\end{figure}

The proposed GRIN metamaterial was composed of 3 layers of unit cells in the direction of
propagation. The corresponding spectral transmission of a 3-layer-structure is presented in
Fig.~\ref{fig:transmission}(b) for different values of the inner slot radius. The structure
exhibits a broad transmission passband for frequencies between $\unit[1.2]{THz}$ and
$\unit[1.9]{THz}$ with an amplitude transmission up to $\unit[85]{\%}$. The enhanced transmission
in the passband is related to the excitation of antisymmetric current modes at the inner and outer
edge of the slot as indicated in Fig.~\ref{fig:unitcell}(a). These antisymmetric modes exhibit a
strongly reduced dipolar coupling to the external field leading to the observed high transmission.

\begin{figure}[t]
   \begin{center}
    \includegraphics[width=7cm]{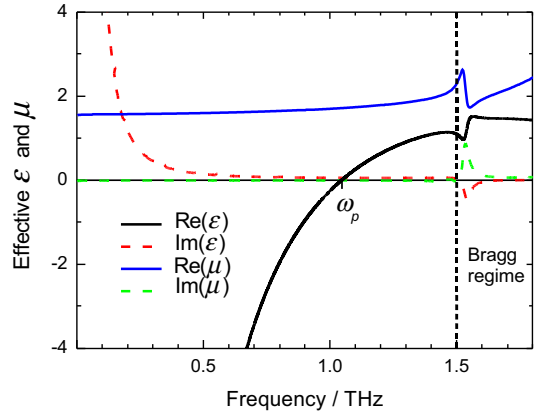}
    \end{center}
    \caption{Retrieved values for the effective permittivity~$\epsilon$ and permeability~$\mu$, exemplarily shown for a slot element with an inner slot radius of $r=\unit[20]{\mbox{\textmu}m}$.}
    \label{fig:epsmue}
\end{figure}

The effective material parameters of the structure were determined by a retrieval algorithm as
described in Ref.~13. The reliability of the effective medium description was corroborated by the
observation that the retrieved parameters were independent of the number of unit cells in the
direction of propagation, i.e.~the thickness of the metamaterial slab. Furthermore, the real part
of the retrieved index of refraction agreed with the values deduced from the phase advance of a
propagating plane wave through the metamaterial slab. In Fig.~\ref{fig:transmission}(a), we present
the retrieved refractive index~$n$ for varying values of the slot radius. Note that for frequencies
above \unit[1.5]{THz}, the half-wavelength inside the medium is smaller than the unit cell size
(Bragg regime) and effective values are not expected to be reliable. The graphs show that the
increasing transmission at the low frequency edge of the passband is related to a rapid increase of
the refractive index from $n\approx0$ to values of about $n\approx1.6-1.8$. This behavior is caused
by a Drude-like response of the effective permittivity whereas the effective permeability is almost
constant (see Fig.~\ref{fig:epsmue}) implying the observed increase of $n$ and of the transmission
for frequencies above the plasma frequency $\omega_\text{p}$. The strong frequency dispersion of
$n$ is a key feature of the proposed GRIN lens since it allows a large variation of $n$ within the
passband by altering the slot radius. As seen from Fig.~\ref{fig:transmission}, the highest index
contrast is achieved in the lower frequency range of the passband at about $\unit[(1.2-1.4)]{THz}$
with values up to $\Delta n = 1.5$. In this frequency interval, the ratio of the vacuum wavelength
to the unit cell size is $\lambda_0/a=3.6-4.2$.\\

\begin{figure}[b]
   \begin{center}
    \includegraphics[width=\columnwidth]{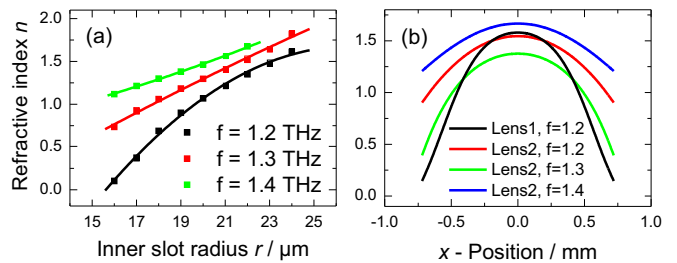}
    \end{center}
    \caption{(a) Retrieved values of the effective index of refraction~$n$ in dependence of the inner radius~$r$ of the annular slot at three different frequencies. (b) Spatial index profiles
    of the proposed highly refractive Lens\,1 and of the broadband Lens\,2 at three different frequencies in units of THz.}
    \label{fig:profile}
\end{figure}

In the following, we illustrate the optical properties of the proposed GRIN metamaterial by two
exemplary realizations: A highly refractive lens with a very large index contrast of $\Delta n =
1.51$ specified at a frequency of \unit[1.2]{THz} (Lens\,1) and a broadband lens applicable in the
range of $\unit[(1.2-1.4)]{THz}$ (Lens\,2). In each case, the GRIN lens was composed as a
circular-shaped lens with a diameter of 25 unit cells, a thickness of 3 unit cells in the direction
of propagation and was framed by a metallic aperture (see Fig.~\ref{fig:unitcell}(b)). The incident
THz wave was simulated by a linearly polarized plane wave and the overall calculation domain was
terminated by open boundary conditions. The spatial refractive index profile of the GRIN lens was
created by varying the inner radius~$r$ of the annular slot elements in dependence on their
position within the lens. The appropriate relation between~$r$ and $n$ was determined from the
retrieved values of~$n$. For example, Fig.~\ref{fig:profile}(a) shows the dependence $n(r)$ at the
frequencies \unit[1.2]{THz}, \unit[1.3]{THz}, and \unit[1.4]{THz}. For the broadband Lens\,2, the
radius was varied between \unit[17]{\mbox{\textmu}m} and \unit[22]{\mbox{\textmu}m}. In this
interval, the relation between~$r$ and~$n$ is almost linear and, thus, the parabolic index profile
was approximated by a parabolic $r$-profile, according to
\begin{align}
r(i,j) = \left(r_{\text{min}}-r_{\text{max}}\right)\frac{i^2+j^2}{N^2} + r_{\text{max}}
\label{eq:profile}
\end{align}
where $i$ and $j$ denote the cell indices in the $x$- and $y$-direction and $N$ is the maximum cell
index. For the highly refractive Lens\,1, the inner radius of the slot elements was varied between
\unit[16]{\mbox{\textmu}m} and \unit[24]{\mbox{\textmu}m}. Here, we optimized the accuracy of the
index profile by determining a higher order fit for the inverse function~$r(n)$ at \unit[1.2]{THz}
and then setting $r(i,j)=r\left(n(i,j)\right)$ where $n(i,j)$ describes a parabolic profile. The
corresponding spatial refractive index profiles of both lenses are plotted in
Fig.~\ref{fig:profile}(b). It should be noted that the index profile of the broadband lens depends
on the considered operation
frequency due to the frequency dispersion of the refractive index.\\


\begin{figure}[t]
   \begin{center}
    \includegraphics[width=\columnwidth]{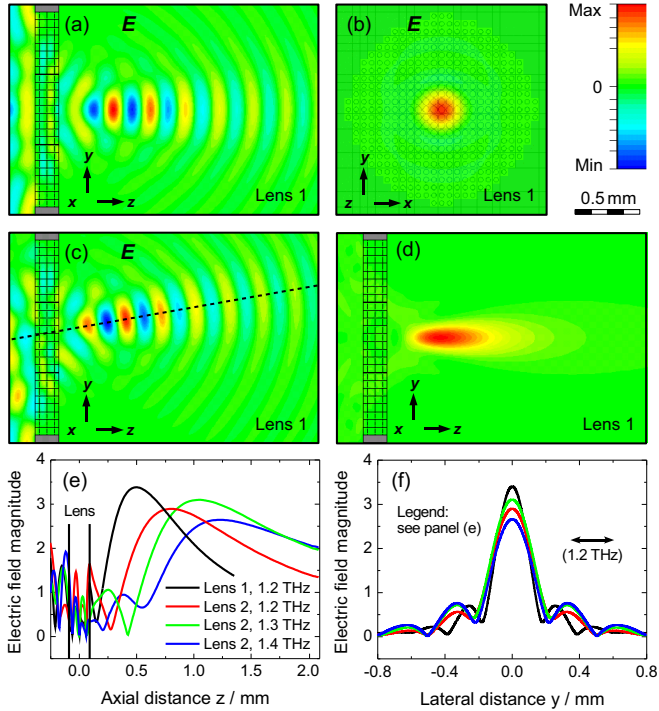}
    \end{center}
    \caption{Results of the full wave simulations of the highly refractive lens (Lens\,1) and the broadband lens (Lens\,2) for an $x$-polarized THz wave incident from the left. (a) $x$-component of the electric field at \unit[1.2]{THz} in the $y$-$z$-plane and (b) in the focal plane for normal incidence (with visible lens
contours indicated in the background). (c)~Same as (a) but for an incidence angle of $\alpha =
\unit[11]{^\circ}$. (d)~Time-averaged intensity in the $y$-$z$-plane. (e)~Magnitude of the electric
field relative to the magnitude of the incident wave along the $z$-axis and (f)~along the $y$-axis
at the focus plane.}
    \label{fig:results}
\end{figure}


In Fig.~\ref{fig:results}, we present the results of the 3D full wave simulations.
Figs.~\ref{fig:results} (a)\,--\,(d) show 2D plots of the electric field on selected planes for the
highly refractive Lens\,1. As expected, the GRIN lens provides the desired focusing functionality
which is evidenced by uniformly curved phase fronts of the electric field in the $y$-$z$-plane
(Fig.~\ref{fig:results}(a)) and the high intensity at the focus (Fig.~\ref{fig:results}(d)).
Moreover, the rotational symmetry of the field distribution in the focal plane shown in
Fig.~\ref{fig:results}(b) indicates that the focusing effect of the GRIN metamaterial is not
sensitive to the polarization of the incident field. Furthermore, the focusing capabilities are
maintained for oblique incidence. This is demonstrated in Fig.~\ref{fig:results}(c) for an incident
wave at an angle of incidence of $\alpha = \unit[11]{^\circ}$. Obviously, the achieved spot size
and focal length are almost identical to those for normal incidence (this holds for both TE and
TM-polarized waves).

For a quantitative analysis of the proposed GRIN lenses, we evaluated the magnitude of the electric
field along the $z$-axis as well as along the $y$-direction at the focal plane for both lens
configurations. The obtained axial and transverse beam profiles, normalized to the magnitude of the
incident THz wave, are presented in Figs.~\ref{fig:results}(e) and \ref{fig:results}(f). The graphs
show that the proposed broadband lens (Lens\,2) provides the desired focusing effect in the entire
spectral range of $\unit[(1.2-1.4)]{THz}$. However, as can be seen from Fig.~\ref{fig:results}(e),
the frequency dependence of the refractive index $n$ results in a noticeable chromatic aberration.
In particular, the focal length ($z$-position where the field magnitude is maximal), increases from
\unit[0.8]{mm} at \unit[1.2]{THz} to \unit[1.2]{mm} at \unit[1.4]{THz}. The increase of the focal
length for higher frequencies agrees with the lowered curvature of the index profile shown in
Fig.~\ref{fig:profile}(b). The achieved focal spot diameters of both lenses are in the order of the
wavelength of the incident THz wave (see Fig.~\ref{fig:results}(f)). As expected, the smallest spot
size was achieved for the highly refractive Lens\,1. According to the common definition of the spot
diameter as the FWHM width of the intensity, the achieved spot diameter for Lens\,1 was
\unit[170]{\mbox{\textmu}m} and, thus, was smaller than the wavelength of the incident THz wave,
$\lambda = \unit[250]{\mbox{\textmu}m}$. Consequently, Lens\,1 also produces the highest intensity
in the focus. As seen from Fig.~\ref{fig:results}(e), the amplitude of the electric field at the
focus is increased by a factor of 3.4 corresponding to an intensity enhancement of more than an
order of
magnitude.\\

In conclusion, we have analyzed and discussed an array of annular slots as a non-resonant, low-loss
gradient index metamaterial. The structure provides a large achievable index contrast up to $\Delta
n = 1.5$ and a sufficient bandwidth for practical applications. The appropriateness of the design
concept was demonstrated for the examples of a highly refractive lens and a broadband lens
operating in the range of $\unit[(1.2-1.4)]{THz}$, both consisting of only 3 layers of unit cells.
The proposed lenses operate irrespectively of the polarization of the incident wave and are
insensitive to the angle of incidence. Furthermore, we showed that the highly refractive lens is
applicable to focus THz radiation to a spot size smaller than the wavelength of the incident wave
where the THz intensity in the focus was increased by more than one order of magnitude. The planar
geometry and the small number of necessary functional layers allow the fabrication of such
metamaterials by standard multilayer BCB lithography techniques as described in Ref.~14. That way,
the proposed design concept is especially advantageous for the development of tailored optical
components for the THz technology.
\cite{chen2004,paul2008}



\begin{thebibliography}{10}
\newcommand{\enquote}[1]{``#1''}
\expandafter\ifx\csname url\endcsname\relax
  \def\url#1{\texttt{#1}}\fi
\expandafter\ifx\csname urlprefix\endcsname\relax\def\urlprefix{URL }\fi
\providecommand{\eprint}[2][]{\url{#2}}

\bibitem{smith2000}
D.~R. Smith, D.~C. Vier, N.~Kroll, and S.~Schultz, \enquote{Direct calculation
  of permeability and permittivity for a left-handed metamaterial,} Appl. Phys.
  Lett. \textbf{77}, 2246--2248 (2000).

\bibitem{liu2009a}
R.~Liu, Q.~Cheng, J.~Y. Chin, J.~J. Mock, T.~J. Cui, and D.~R. Smith,
  \enquote{Broadband gradient index microwave quasi-optical elements based on
  non-resonant metamaterials,} Opt. Express \textbf{17}, 21,030--21,041 (2009).

\bibitem{cheng2009}
Q.~Cheng, H.~F. Ma, and T.~J. Cui, \enquote{Broadband planar Luneburg lens
  based on complementary metamaterials,} App. Phys. Lett. \textbf{95}, 181,901
  (2009).

\bibitem{liu2009b}
R.~Liu, C.~Ji, J.~J. Mock, J.~Y. Chin, T.~J. Cui, and D.~R. Smith,
  \enquote{Broadband Ground-Plane Cloak,} Science \textbf{323}, 366 (2009).

\bibitem{cai2007}
W.~Cai, U.~K. Chettiar, A.~V. Kildishev, and V.~M. Shalaev, \enquote{Optical
  cloaking with metamaterials,} Nature Photon. \textbf{1}, 224 -- 227 (2007).

\bibitem{ruan2007}
Z.~Ruan, M.~Yan, C.~W. Neff, and M.~Qiu, \enquote{Ideal Cylindrical Cloak:
  Perfect but Sensitive to Tiny Perturbations,} Phys. Rev. Lett. \textbf{99},
  113,903 (2007).

\bibitem{pendry2008}
J.~Li and J.~B. Pendry, \enquote{Hiding under the Carpet: A New Strategy for
  Cloaking,} Phys. Rev. Lett. \textbf{101}, 203,901 (2008).

\bibitem{leonhardt2006}
U.~Leonhardt, \enquote{Optical Conformal Mapping,} Science \textbf{312},
  1777--1780 (2006).

\bibitem{pendry2006}
J.~B. Pendry, D.~Schurig, and D.~R. Smith, \enquote{Controlling Electromagnetic
  Fields,} Science \textbf{312}, 1780--1782 (2006).

\bibitem{schurig2007}
D.~Schurig, J.~B. Pendry, and D.~R. Smith, \enquote{Transformation-designed
  optical elements,} Opt. Express \textbf{15}, 14,772 (2007).

\bibitem{rahm2008}
M.~Rahm, S.~A. Cummer, D.~Schurig, J.~B. Pendry, and D.~R. Smith,
  \enquote{Optical Design of Reflectionless Complex Media by Finite Embedded
  Coordinate Transformations,} Phys. Rev. Lett. \textbf{100}, 063,903 (2008).

\bibitem{kundtz2010}
N.~Kundtz and D.~R. Smith, \enquote{Extreme-angle broadband metamaterial lens,}
  Nat. Mater. \textbf{9}, 129–132 (2010).

\bibitem{chen2004}
X.~Chen, T.~M. Grzegorczyk, B.-I. Wu, J.~{Pacheco, Jr.}, and J.~A. Kong,
  \enquote{Robust method to retrieve the constitutive effective parameters of
  metamaterials,} Phys. Rev. E \textbf{70}, 016,608 (2004).

\bibitem{paul2008}
O.~Paul, C.~Imhof, B.~Reinhard, R.~Zengerle, and R.~Beigang, \enquote{Negative
  index bulk metamaterial at terahertz frequencies,} Opt. Express
  \textbf{16}(9), 6736--6744 (2008).

\end{thebibliography}

\end{document}